\documentclass{elsart}
\usepackage{amssymb,amsmath}
\usepackage {graphicx}
\usepackage[T2A]{fontenc}
\usepackage[english]{babel}
\usepackage{amssymb}

\begin{document}

\begin{frontmatter}



\title{
A non-autonomous flow system \\ with Plykin type attractor }


\author{Sergey P. Kuznetsov}

\address{Kotel'nikov Institute of Radio-Engineering and Electronics 
of RAS,\\ 
Saratov Branch, Zelenaya 38, 410019, Saratov, Russian Federation}

\begin{abstract}
A non-autonomous flow system is introduced with an attractor of Plykin type 
that may serve as a base for elaboration of real systems and devices 
demonstrating the structurally stable chaotic dynamics. The starting point 
is a map on a two-dimensional sphere, consisting of four  
stages of continuous geometrically evident transformations. The computations 
indicate that in a certain parameter range the map has a uniformly 
hyperbolic attractor. It may be represented on a plane by means of a 
stereographic projection. Accounting structural stability, a modification of 
the model is undertaken to obtain a set of two non-autonomous differential 
equations of the first order with smooth coefficients. 
As follows from computations,  
it has the Plykin type attractor in the
Poincar\'{e} cross-section. 
\end{abstract}

\end{frontmatter}

In mathematical theory of dynamical systems a class of uniformly hyperbolic 
chaotic attractors is known \cite{01,02,03,04,05,06,07}. In such an attractor all orbits are of 
the saddle type, and their stable and unstable manifolds do not touch each 
other, but can only intersect transversally. These attractors manifest 
strong stochastic properties and allow detailed mathematical analysis. They 
are structurally stable; that means insensitivity of the structure of the 
attractors in respect to variation of functions and parameters in the 
dynamical equations. Although main concepts of the relevant mathematical 
theory were advanced 40 years ago, until recently, these attractors are 
considered rather as a refined image of chaos, than as adequate models for 
real-world systems. In textbooks and reviews examples of the uniformly 
hyperbolic attractors traditionally are represented by mathematical 
constructions, like Plykin attractor and Smale -- Williams solenoid. These 
examples relate to discrete-time systems, iterated maps. In particular, 
Plykin attractor takes place in some special map on a sphere with four 
holes, or on a plane in a bounded domain with three holes \cite{06}. In 
applications, physics and technology people deal usually with systems 
operating in continuous time, called flows in mathematical literature. In 
such systems Plykin type attractors could occur in the context of 
description based on the Poincar\'{e} map \cite{07,08,09}.

The present work is aimed at explicit construction of a non-autonomous flow 
system with Plykin type attractor, which could provide a basis for  
development of real systems and devices, demonstrating structurally stable 
chaotic dynamics. The starting point is consideration of
a motion on a 
two-dimensional sphere composed of periodically repeated stages 
of continuous geometrically evident transformations.

Let us consider a sphere of unit radius. A point on the sphere can be 
specified in angular coordinates $(\theta ,\,\varphi )$, or in Cartesian 
coordinates
\begin{equation}
\label{eq1}
x = \cos \varphi \sin \theta ,
\quad
y = \,\sin \varphi \sin \theta ,
\quad
z = \cos \theta ,
\end{equation}
which satisfy the relation $x^2 + y^2 + z^2 = 1$. As proven by Plykin, a map 
on a sphere can possess hyperbolic attractor only in the presence of at 
least four holes, the areas not visited by trajectories belonging to the 
attractor. In our construction this role will be played by neighborhoods of 
four points A, B, C, D with coordinates $(x,y,z) = (\pm 1 / \sqrt 2 
,\,\,0,\,\,\pm 1 / \sqrt 2 )$. 

Let us consider a sequence of four successive continuous transformations, 
each of which is of unit time duration. 

\textbf{I. Flow down along circles of latitude,} that is motion of the 
representative points on the sphere away from the meridians NABS and NDCS
(N and S designate the north and the south poles) 
towards the meridians equally distant from the arcs AB and CD. In Cartesian 
coordinates it is governed by equations 
\begin{equation}
\label{eq2}
\dot {x} = - \varepsilon xy^2,\,\,\,\dot {y} = \varepsilon x^2y,\,\,\,\,\dot 
{z} = 0,
\end{equation}
where $\varepsilon $ is a parameter.

\textbf{II. Differential rotation} around $z$-axis with angular velocity 
depending on $z$ linearly, in such way that the points B and C do not move, 
while A and D exchange their location; it corresponds to equations
\begin{equation}
\label{eq3}
\dot {x} = \pi (z / \sqrt 2 + 1 / 2)y,\,\,\,\,\dot {y} = - \pi (z / \sqrt 2 
+ 1 / 2)x,\,\,\,\,\dot {z} = 0.
\end{equation}

\textbf{III. Flow down to the equator}, that is motion of representative 
points along circles centered on the $x$-axis on the sphere from the great 
circle ABCD, towards the equator:
\begin{equation}
\label{eq4}
\dot {x} = 0,\,\,\,\dot {y} = \varepsilon yz^2,\,\,\,\dot {z} = - 
\varepsilon y^2z.
\end{equation}

\textbf{IV. Differential rotation} around $x$-axis with angular velocity 
depending on $x$ linearly, in such way that representative points 
in the plane, orthogonal to the $ x$-axis and containing the point C, 
do not move, 
while those in 
the plane containing the point B undergo a turn by
$180^{\circ}$:
\begin{equation}
\label{eq5}
\dot {x} = 0,\,\,\,\dot {y} = - \pi (x / \sqrt 2 + 1 / 2)z,\,\,\,\dot {z} = 
\pi (x / \sqrt 2 + 1 / 2)y.
\end{equation}

Note a symmetry of the procedure: the first and the second pairs of the 
transformations are identical, up to exchange of the variables $x$ and $z$. 

The Poincar\'{e} map describing transformation of a state vector ${\rm {\bf 
x}}_n = (x_n ,y_n ,z_n )$ on a period $T = 4$ may be obtained explicitly. 
From successive solution of the differential equations (\ref{eq2})-(\ref{eq5}) with account 
of the mentioned symmetry, one can represent the resulting state vector 
${\rm {\bf x}}_{n + 1} $ as 
\begin{equation}
\label{eq6}
{\rm {\bf x}}_{n + 1} = {\rm {\bf f}}_{+} ({\rm {\bf f}}_{ - } ({\rm {\bf 
x}}_n )),\,\,
{\rm {\bf f}}_{\pm} ({\rm {\bf x}}) = \left\{ 
{{\begin{array}{*{20}c}
 {\pm z,} \hfill \\
\frac{\sqrt{x^2 + y^2 }
\left (
y e^{ \varepsilon (x^2 + y^2 )/2} 
\cos {\pi \over 2}(z \sqrt 2 + 1) \pm 
x e^{-\varepsilon (x^2 + y^2 )/2}
\sin {\pi \over 2}(z \sqrt 2 + 1)
\right )}
{\sqrt {x^2 
e^{- \varepsilon (x^2 + y^2 )} + 
y^2 e^{\varepsilon (x^2 + y^2 )}}},
 \\
\frac{\sqrt{x^2 + y^2 }
\left (
y e^{ \varepsilon (x^2 + y^2 )/2} 
\sin {\pi \over 2}(z \sqrt 2 + 1) \mp 
x e^{-\varepsilon (x^2 + y^2 )/2}
\cos {\pi \over 2}(z \sqrt 2 + 1)
\right )}
{\sqrt {x^2 
e^{- \varepsilon (x^2 + y^2 )} + 
y^2 e^{\varepsilon (x^2 + y^2 )}}}.
 \\
\end{array} }} \right.
\end{equation}

The relations (\ref{eq6}) determine the map on the sphere ${\rm {\bf 
x}}_{n + 1} = {\rm {\bf T}}({\rm {\bf x}}_n )$. Note that $C$ is a fixed point 
of the map ${\rm {\bf T}}$, while $A$, $B$ and $D$ compose an unstable periodic orbit 
of period 3: $A \to D \to B \to A$. The map ${\rm {\bf T}}$ is invertible. 
The inverse map appears as a result of the same transformations in backward 
order, with reversed directions of the rotations. 

Figure 1 shows attractor of the map \textbf{T} at $\varepsilon $=0.77. 
Observe specific fractal-like transversal structure of the attractor: the 
object looks like composed of strips, each of which contains narrower strips 
of the next level etc. 

Description of the dynamics can be reformulated to represent instantaneous 
states of the system on a plane. The variable change is 
\begin{equation}
\label{eq7}
W = X + iY = \frac{x - z + iy\sqrt 2 }{x + z + \sqrt 2 },
\end{equation}
which corresponds to a stereographic projection, with selection of the 
projection point at $C ( - 1 / \sqrt 2 ,\,\,0,\,\, - 1 / \sqrt 2 )$. This 
point does not belong to the attractor (it is in the ``hole''), so the image 
of the attractor on the plane is located in a bounded domain. Portrait of 
the attractor for the Poincar\'{e} map in this representation is shown in 
Fig.2a.
\begin{figure}[htbp]
\centerline{\includegraphics[width=2.4in]{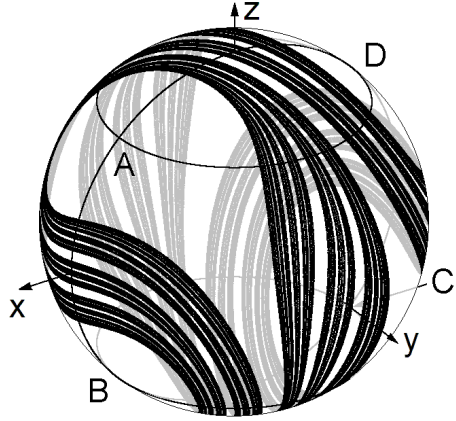}}
\label{fig1}
\caption{
Attractor of the map (\ref{eq6}) at $\varepsilon $=0.77 on the unit 
square
}
\end{figure}

\begin{figure}[htbp]
\centerline{\includegraphics[width=5.6in]{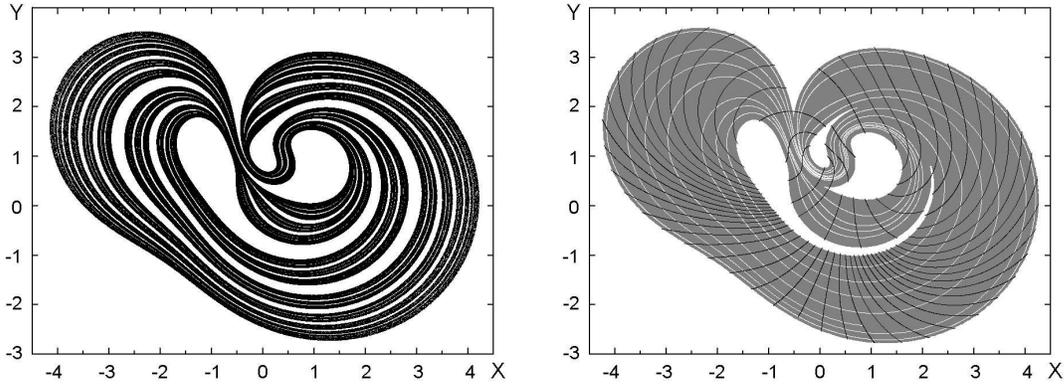}}
\label{fig2}
\caption{
(a) Portrait of the attractor of the map (\ref{eq6}) at $\varepsilon 
$=0.77 presented on a plane by means of the stereographic projection (\ref{eq7}) and 
(b) disposition of stable (black) and unstable (white) manifolds in a domain 
$D$ (gray background) that contains the attractor
}
\end{figure}

I argue that the attractor relates to the uniformly hyperbolic class, i.e. 
it is an attractor of Plykin type. To support this assertion let us turn to 
Fig.~2b illustrating location of stable and unstable manifolds in a domain D 
that contains the attractor.
\footnote{ To draw the stable and unstable 
manifolds with a computer, we do the following. First, for a given 
point on the attractor ${\rm {\bf x}}$ we obtain its image by iterations of 
the Poincar\'e map ${\rm {\bf \bar {x}}} = T_P^N ({\rm {\bf x}})$, and its 
pre-image by iterations of the inverse map 
${\rm {\bf \tilde {x}}} = T_P^{-N} ({\rm {\bf x}})$, 
where $N$ is some empirically chosen integer. Then, with 
random initial conditions ${\rm {\bf \tilde {y}}}$ in a small neighborhood of 
${\rm {\bf \tilde {x}}}$, we get a set of 
points ${\rm {\bf y}} = T_P^N ({\rm {\bf \tilde {y}}})$ 
by iterations of the Poincar\'e map, which mark the 
unstable manifold. In a similar way, starting with random initial conditions 
${\rm {\bf \bar {y}}}$ in a small neighborhood of ${\rm {\bf \bar {x}}}$, we 
draw the stable manifold with a set of points ${\rm {\bf y}} = T_P^{ - N} 
({\rm {\bf \bar {y}}})$.The accuracy the manifolds are depicted grows fast 
with $N$. Actually, $N$ = 6 is enough to get so 
small errors that they are indistinguishable in the plot.\par } 
The 
stable and unstable manifolds are shown, respectively, by black an white 
curves, drawn on a gray background of the domain D. As seen, the 
unstable manifolds follow along filaments of the attractor, while the stable 
ones are transversal to the filaments. Mutual disposition of the stable and 
unstable manifolds certainly excludes possibility of tangencies, at least in 
the domain D.

An alternative approach to verification of the hyperbolicity can be based on 
the cone criterion known from the mathematical literature 
\cite{01,02,03,04,05,09}. Such 
calculations were performed on a base of methodology developed in 
Ref.\cite{10}, 
and the hyperbolicity was confirmed.\footnote{These results will be 
published elsewhere, because require more volume than 
appropriate for the short communication.}

Lyapunov exponents were computed for the map ${\rm {\bf T}}$ by means of the 
procedure based on joint iterations of the map together with a collection of 
two sets of linearized equations for perturbation vectors. At each 
iteration step, these two vectors are orthogonalized, and normalized to a 
fixed constant. Lyapunov exponents are obtained as slopes of the straight 
lines approximating the accumulating sums of logarithms of the norm ratios 
for the vectors in dependence of the number of iterations \cite{11}. In 
particular, at $\varepsilon $=0.77 the Lyapunov exponents are $\Lambda _1 = 
0.9587$ and $\Lambda _2 = - 1.1406$. Then, the estimate of the attractor 
dimension from the Kaplan -- Yorke formula yields $D_{KY} \approx 1 + 
\Lambda _1 / \vert \Lambda _2 \vert = 0.8405$.

One can accept an interpretation that an instant speed of a representative 
point on the sphere is determined by combination of two vector fields, which 
are switched on and off, turn by turn. One corresponds to dynamics during 
the stages of flow down, and another to the stages of differential rotation. 
Let us introduce angles $\bar {\alpha }$ and $\bar {\beta }$ determining 
directions of the fields, and coefficients $p$ and $q$ responsible for switching 
them on and off. We set $\bar {\alpha } = \bar {\beta } = \textstyle{\pi 
\over 4}( - 1)^{[t / 2]} - \textstyle{\pi \over 4},\,\,p = \textstyle{1 
\over 2} + \textstyle{1 \over 2}( - 1)^{[t]},\,\,\,\,q = \textstyle{1 \over 
2}( - 1)^{[t / 2]} - \textstyle{1 \over 2}( - 1)^{[t / 2 + 1 / 2]}$. (Here 
[$\tau $] designates the integer part of $\tau $.) Setting $K = \pi / \sqrt 
2 $, we write down the Eqs.~(\ref{eq2})-(\ref{eq5}) in a compact form:
\begin{equation}
\label{eq8}
\begin{array}{l}
 \dot {x} = - p\varepsilon y^2(x\cos \bar {\alpha } + z\sin \bar {\alpha 
})\cos \bar {\alpha } + qKy( - x\sin \bar {\beta } + z\cos \bar {\beta } + 
\textstyle{1 \over {\sqrt 2 }})\cos \bar {\beta },\,\,\, \\ 
 \dot {y} = p\varepsilon y(x\cos \bar {\alpha } + z\sin \bar {\alpha })^2 - 
qK(x\cos \bar {\beta } + z\sin \bar {\beta })( - x\sin \bar {\beta } + z\cos 
\bar {\beta } + \textstyle{1 \over {\sqrt 2 }}),\,\,\, \\ 
 \dot {z} = - p\varepsilon y^2(x\cos \bar {\alpha } + z\sin \bar {\alpha 
})\sin \bar {\alpha } + qKy( - x\sin \bar {\beta } + z\cos \bar {\beta } + 
\textstyle{1 \over {\sqrt 2 }})\sin \bar {\beta }. \\ 
 \end{array}
\end{equation}

As a next step, let us construct a version of the non-autonomous model 
containing only smooth functions. To do this, in Eqs.~(\ref{eq8}) we simply set 
\begin{equation}
\label{eq9}
\alpha = \textstyle{\pi \over 4}\cos \textstyle{\pi \over 2}t,\,\,\beta = 
\textstyle{\pi \over 4}\sin \textstyle{\pi \over 2}t,\,\,p = 1,\,\,\,q = 
\sin \textstyle{\pi \over 2}t.
\end{equation}

Now, two vector fields, responsible for the flow down and for the 
differential rotation vary in time continuously and smoothly, undergoing 
rotations in space in such way that the original configuration is repeated 
with the period $T = 4$. The second field oscillates, reversing the 
direction twice on a period. 
The time variations of one and other fields are 
shifted in phase relatively by a quarter of period. So, the extremal values of 
the fields are achieved turn by turn. The action of the fields at the 
extrema are mostly significant for motion of the representative points, 
hence, it is reasonable to specify the values at extrema, 
like in the original model. Because of structural 
stability of the hyperbolic attractor, one can hope that its nature remains 
the same after the modification, at least in a properly selected range of 
the parameters. As seen from computations, it is so, e.g. at 
$\varepsilon $=0.72 and $K$=1.9.

In contrast to the previous version of the model, the Poincar\'{e} map can 
not be expressed analytically. However, it may be easily realized by a 
computer program integrating the equations with a finite-difference method 
on a time period $T$ = 4.

We can exclude a redundant variable in the equations by means of the 
variable change (\ref{eq7}), and rewrite them in terms of $W = X + iY$. After 
separation of the real and imaginary parts, we obtain
\begin{equation}
\label{eq10}
\begin{array}{c}
 \dot{X} = - 2\varepsilon Y^2 \Omega_1 (X,Y,t)\left( {\cos 
(\textstyle{\pi \over 4}\cos \textstyle{\pi \over 2}t) - X\sin 
(\textstyle{\pi \over 4}\cos \textstyle{\pi \over 2}t)} \right) \\ 
+ KY\Omega _2 (X,Y,t)\left( {\cos 
(\textstyle{\pi \over 4}\sin \textstyle{\pi \over 2}t) - X\sin 
(\textstyle{\pi \over 4}\sin \textstyle{\pi \over 2}t)} \right)\,\sin 
\textstyle{{\pi t} \over 2}, \\
\\ 
\dot{Y} = \varepsilon Y\Omega _1 (X,Y,t)\left( {2X\cos 
(\textstyle{\pi \over 4}\cos \textstyle{\pi \over 2}t) + 
(1 - X^2 + Y^2)\sin (\textstyle{\pi \over 4}\cos \textstyle{\pi \over 
2}t)} \right) \\ 
- K\Omega _2 (X,Y,t)\left( {X\cos 
(\textstyle{\pi \over 4}\sin \textstyle{\pi \over 2}t) + \textstyle{1 \over 
2}(1 - X^2 + Y^2)\sin (\textstyle{\pi \over 4}\sin \textstyle{\pi \over 
2}t)} \right)\,\sin \textstyle{{\pi t} \over 2}, \\ 
 \end{array}
\end{equation}
where
\begin{equation}
\label{eq11}
\begin{array}{c}
\Omega _1 (X,Y,t) = 
\frac{2X\cos({\pi \over 4}\cos{\pi \over 2}t)
+(1-X^2-Y^2)\sin({\pi \over 4}\cos{\pi \over 2}t)
}{(1+X^2+Y^2)^2}, \\ 
\\
\Omega _2 (X,Y,t) = 
\frac{-2X\sin({\pi \over 4}\sin{\pi \over 2}t)
+(1-X^2-Y^2)\cos({\pi \over 4}\sin{\pi \over 2}t)
}{(1+X^2+Y^2)}+\frac{1}{\sqrt 2}. \\ 
 \end{array}
\end{equation}

The expressions (\ref{eq10}), (\ref{eq11}) look a bit unwieldy, 
but this is, apparently, 
the first explicit example of a set of differential equations 
with smooth coefficients, which has attractor of Plykin type 
on a plane in the Poincar\'{e} cross-section.
\begin{figure}[b]
\centerline{\includegraphics[width=5.6in]{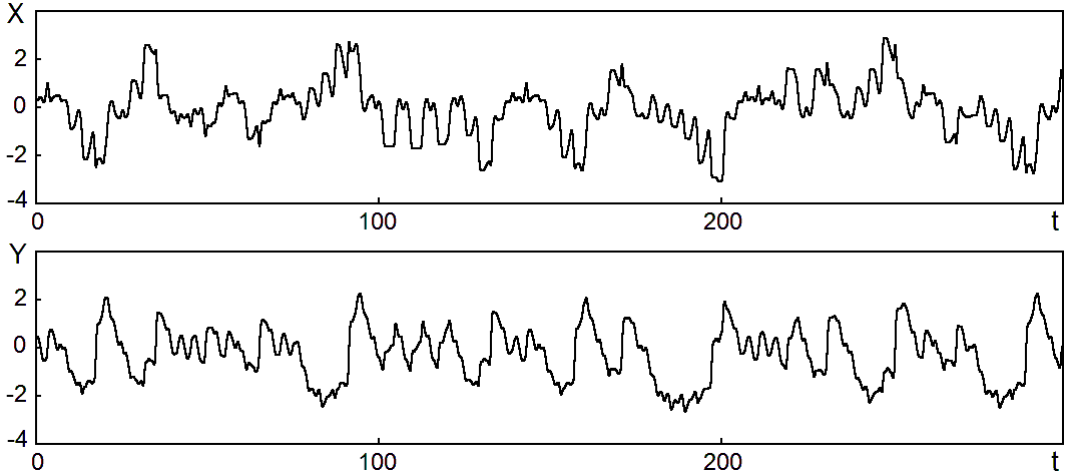}}
\label{fig3}
\caption{
Variables $X$ and $Y$ versus time as obtained from numerical
integration of the differential equations (\ref{eq10}) at $\varepsilon $=0.72 and 
$K$=1.9. The plot relates to sustained chaotic regime associated with motion on 
the attractor; the transients are excluded
}
\end{figure}

Figure~3 shows variables $X$ and $Y$ in dependence on time as obtained from 
numerical integration of the differential equations (\ref{eq10}) at $\varepsilon 
$=0.72 and $K$=1.9, after exclusion of transients. Visually they look like 
realizations of random processes, as it should be for dynamics on the 
chaotic attractor.

\begin{figure}[htbp]
\centerline{\includegraphics[width=3in]{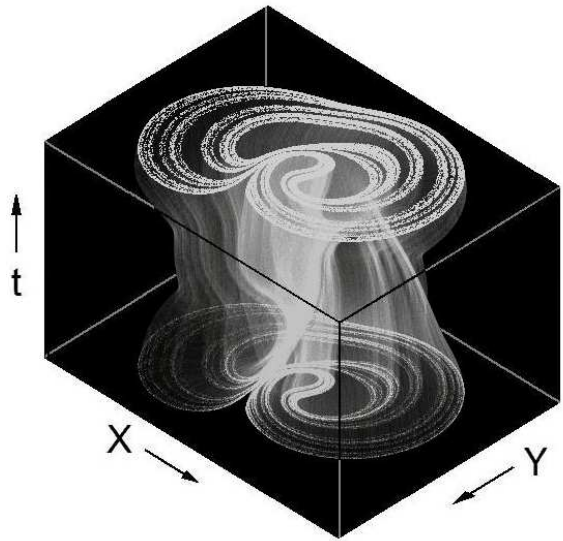}}
\label{fig4}
\caption{
Portrait of attractor of the model (\ref{eq10}) at $\varepsilon 
$=0.72 and $K$=1.9 in the extended three-dimensional phase space. The 
gray-scale technique is used: brighter tones correspond to  
relatively larger probability of visiting the pixels 
by orbits on the attractor}
\end{figure} 

Figure 4 shows portrait of the attractor in the three-dimensional extended 
phase space ($X$,~$Y$,~$t)$. To make visible the inherent structure, 
the picture is 
presented in the gray-scale technique. Brighter tones correspond to 
relatively larger probability of visiting pixels by orbits on the attractor. 
Time interval on the vertical axis corresponds just to a period of variation 
of coefficients in the equations. In the cross-section of the attractor with 
the horizontal plane, one can observe an object with fractal structure, 
remarkably similar to that discussed in the context of the sphere map (see 
Fig. 2a). At a qualitative level, it may be regarded as an argument in favor 
of persistence of the hyperbolic attractor under modification of the model 
we undertake. 

Computation of the Lyapunov exponents for the model (\ref{eq10}) by means of the 
Benettin algorithm \cite{11} at $\varepsilon $=0.72 and $K$=1.9 yields $\lambda _1 
\approx 0.221$ and $\lambda _2 \approx - 0.315$. It corresponds to the 
Lyapunov exponents of the Poincar\'{e} map $\Lambda _1 = \lambda _1 T 
\approx 0.884$ and $\Lambda _2 = \lambda _2 T \approx - 1.260$. Estimate of 
the attractor dimension in the Poincar\'{e} section by the Kaplan -- Yorke 
formula is $D_{KY} \approx 1.70$.

The hyperbolic nature of the attractor was verified by means of graphical 
representation of manifolds in the Poincar\'{e} section (like in Fig.2) and 
by the computations based on the cone criterion (in analogy 
with Ref.~\cite{11}). 
As follows from those results, attractor is hyperbolic in some parameter 
range around $\varepsilon $=0.72 and $K$=1.9. (More details will be published 
elsewhere.)

To conclude, this work puts into consideration a non-autonomous flow system manifesting 
chaotic dynamics associated with a hyperbolic strange attractor. In the 
stroboscopic Poincar\'{e} map it is attractor of Plykin type. In fact, 
I present
two versions of the model. In the first version the evolution
consists of successive stages, and the differential equations have piecewise
continuous dependency of coefficients on time. The 
Poincar\'{e} map is expressed analytically, as a map on a sphere 
In the second version, the system is modified in such way that 
it is governed by a set of differential equations with smooth 
coefficients. The modification does not alter the hyperbolic nature of the 
chaotic attractor due to structural stability intrinsic to this object.
As the system has the minimal phase space dimension required for 
existence of a hyperbolic strange attractor, its investigation, including 
verification of the hyperbolicity criteria is essentially simpler, in 
comparison with models suggested earlier 
as examples of attractors of Smale-Williams type \cite{10,13,14}. 
Appearance of concrete examples of systems with hyperbolic strange 
attractors is of evident significance both from the point of view of 
complementation of mathematical concepts with concrete and visible context 
(see e.g. Ref.~\cite{12}), and for exploiting these concepts in applications. 
Hyperbolic chaotic systems may be of special interest for applications 
due to structural stability, that means insensitivity of the generated 
chaos to variations of parameters, characteristics of elements, technical 
fluctuations etc.

\textit{The work was performed, in part, during a visit of the author 
to the Group of Statistical Physics and Theory of Chaos in Potsdam 
University. The research is supported by RFBR-DFG grant 08-02-91963 
and by grant 2.1.1/1738 of Ministry of Education and Science of 
Russian Federation in a frame of program of Development of 
Scientific Potential of Higher Education}.

\end{document}